\documentclass[reprint,superscriptaddress,amsmath,amssymb,aps,prb,longbibliography]{revtex4-2}
\usepackage[utf8]{inputenc}
\usepackage{graphicx}
\usepackage{dcolumn}
\usepackage{bm}
\usepackage[colorlinks,citecolor=blue,linkcolor=blue,urlcolor=blue]{hyperref}
\usepackage[normalem]{ulem}
\usepackage{newunicodechar}
\newunicodechar{（}{(}
\newunicodechar{）}{)}

\begin{document}

\title{Variation of Bose surface by Filling in Cooper pair Bose metal}

\author{Jiahao Su}
\affiliation{School of Science, Harbin Institute of Technology, Shenzhen, 518055, China}
\affiliation{Shenzhen Key Laboratory of Advanced Functional Carbon Materials Research and Comprehensive Application, Shenzhen 518055, China.}

\author{Ji Liu}
\affiliation{School of Science, Harbin Institute of Technology, Shenzhen, 518055, China}
\affiliation{Shenzhen Key Laboratory of Advanced Functional Carbon Materials Research and Comprehensive Application, Shenzhen 518055, China.}

\author{Jianyu Li}
\affiliation{School of Science, Harbin Institute of Technology, Shenzhen, 518055, China}
\affiliation{Shenzhen Key Laboratory of Advanced Functional Carbon Materials Research and Comprehensive Application, Shenzhen 518055, China.}

\author{Zhangkai Cao}
\email{caozhangkai@mail.bnu.edu.cn}
\affiliation{Eastern Institute for Advanced Study, Eastern Institute of Technology, Ningbo, Zhejiang 315200, China}
\affiliation{School of Physical Sciences, University of Science and Technology of China, Hefei, 230026, China}

\author{Tao Ying}
\affiliation{School of Physics, Harbin Institute of Technology, Harbin 150001, China}

\author{Ho-Kin Tang}
\email{denghaojian@hit.edu.cn}
\affiliation{School of Science, Harbin Institute of Technology, Shenzhen, 518055, China}
\affiliation{Shenzhen Key Laboratory of Advanced Functional Carbon Materials Research and Comprehensive Application, Shenzhen 518055, China.}

\date{\today}

\begin{abstract}

The Cooper pair Bose metal (CPBM) is a non-superfluid quantum phase in which uncondensed fermion pairs form a ``Bose surface (BS)” in momentum space. We investigate the CPBM in the two-dimensional spin-anisotropic attractive Hubbard model by tuning the next-nearest-neighbor (NNN) hopping $t^{\prime}$, carrier filling $n$, and spin anisotropy $\alpha$, using large-scale constrained-path quantum Monte Carlo simulations. A moderate NNN hopping $t^{\prime}/t = 0.2$ substantially enlarges the CPBM region: the phase extends into weaker anisotropy regimes and coexists with a commensurate charge-density wave (CDW) near half-filling ($n > 0.95$), where CDW order would otherwise dominate at $t^{\prime} = 0$. Interestingly, $t^{\prime}$ suppresses the overall CDW peak amplitude and introduces a geometric correlation between the orientations of the Fermi and Bose surfaces: for weak Fermi-surface rotations, the Bose surface remains aligned with the lattice axes, while larger distortions drive both surfaces to rotate in tandem. Momentum-resolved pairing distributions reveal that the bosonic pairing channels are jointly controlled by $t^{\prime}$ and carrier filling $n$. For small $t^{\prime}$, $d_{xy}$-wave correlations dominate across the entire filling range. In contrast, for larger $t^{\prime}$, the dominant pairing symmetry varies with $n$, reflecting a nontrivial interplay between frustration and density. These findings establish carrier filling and NNN hopping as complementary levers for manipulating CPBM stability and provide concrete criteria for identifying non-superfluid bosonic matter in cold-atom and correlated-electron systems.

\end{abstract}
\pacs{Valid PACS appear here}

\maketitle

\section{Introduction}
Fermi-liquid (FL) theory has long served as the foundation of our understanding of conventional metals, successfully describing their ground states and low-energy excitations through the framework of long-lived quasiparticles~\cite{Landau1957,Anderson1998,PinesNozieres1966}. Nevertheless, recent experimental observations in strongly correlated systems have increasingly challenged the adequacy of FL theory. Notable deviations include linear-in-temperature resistivity, strange-metal behavior under pressure, non-Fermi-liquid spin response, and quantum critical scaling near metal-insulator transitions~\cite{Abbamonte2022,Si2012,Anderson1987,Li2024,Sun2023,Wang2024,Phillips2022}. These anomalies are particularly evident in high-$T_c$ cuprates, where the critical temperature surpasses the BCS limit and pseudogap behavior challenges conventional pairing mechanisms~\cite{Keimer2015,Hashimoto2014,RevModPhys.96.025002}. Such observations underscore the urgent need for new theoretical paradigms capable of capturing collective behavior beyond the quasiparticle picture.

One promising direction is the study of Bose metals (BMs)—gapless, non-superfluid quantum phases governed by bosonic excitations such as spinons, or vortices~\cite{RevModPhys.96.025002,Block2011}. In contrast to FLs, BMs exhibit finite conductivity without phase coherence, suggesting a coexistence of strong quantum fluctuations and bosonic transport. In two-dimensional superconducting films, BMs behavior has been experimentally identified as an intermediate phase in the superconductor–insulator quantum phase transition, with disorder and vortex dynamics playing key roles~\cite{Das2001,Shin2025,Phillips2003}. Notably, in perforated Ta films, vortex pinning was shown to suppress superconductivity and stabilize a metallic state~\cite{Zhang2019}, providing a concrete platform for testing bosonic transport models.

A theoretically cleaner realization is the Cooper pair Bose metal (CPBM), where strongly interacting fermions form preformed, uncondensed Cooper pairs that generate a gapless “Bose surface (BS)” in momentum space~\cite{Phillips2003,PhysRevB.78.054520}. This phase lies outside both Fermi-liquid and Bose-condensate categories and has garnered growing attention for its potential to explain the anomalous metallic phases in cuprates and other strongly correlated materials. Early theoretical studies proposed that spin-dependent Fermi surface (FS) anisotropy and frustrated ring-exchange interactions in quasi-one-dimensional geometries (e.g., two- and four-leg ladders) could stabilize CPBM states with $d$-wave symmetry~\cite{Feiguin2009,Feiguin2011}. However, extending these insights to two-dimensional (2D) lattices raised critical concerns: could CPBM phases survive in the presence of strong quantum fluctuations and competing orders, such as charge-density wave (CDW) and conventional $s$-wave superfluidity (s-SF)?

A breakthrough was achieved through recent numerical advances. CPQMC simulations and FRG analysis collectively demonstrated that a moderate spin-dependent nearest-neighbor hopping anisotropy ($\alpha \sim 0.2$) can suppress CDW order and stabilize a $d_{xy}$-wave CPBM phase in the 2D spin-anisotropic attractive Hubbard model\cite{Cao2024}. Subsequent work demonstrated that next-nearest-neighbor (NNN) hopping $t^{\prime}$ further expands the CPBM regime via geometric frustration, while preserving the $d_{xy}$-wave pairing symmetry for $t^{\prime}/t \sim 0.2$~\cite{Cao2025}. Remarkably, larger $t^{\prime}$ ($\gtrsim 0.7$) induces a transition to dominant $d_{x^2 - y^2}$-wave pairing, reminiscent of pseudogap phenomenology in cuprates. These results indicate that the CPBM phase may encode universal features of correlated electron systems and persist even when the FS transitions from open to closed topology—challenging the belief that quasi-1D nesting is a prerequisite for Bose-metal formation.

Despite these advances, key questions remain unresolved. The precise role of carrier density $n$ at fixed $t^{\prime}/t = 0.2$, the parameter regime where the CPBM is most stable, remains to be clarified. The evolution of FS geometry and its rotation with filling, as well as whether the BS consistently tracks these changes, requires further investigation. It is also necessary to determine whether the CPBM can coexist with CDW order near half-filling. Furthermore, a comprehensive understanding of the dominant internal pairing symmetries across the full parameter space is still lacking.

In this work, we address these questions by performing high-precision CPQMC simulations on the 2D spin-anisotropic attractive Hubbard model at fixed $t^{\prime}/t = 0.2$. Our key findings are as follows. First, we demonstrate that the CPBM phase expands significantly and persists up to half-filling, where it coexists with weak commensurate CDW order—revealing a previously unexplored competition regime. Second, we uncover a geometric coupling mechanism whereby the BS remains locked to lattice directions for small FS rotations, but gradually co-rotates with the FS when the Fermi surface undergoes substantial rotation. Finally, we analyze the internal pairing structure and uncover a structure among bosonic pairing channels that is jointly shaped by $t^{\prime}$ and filling $n$. At small $t^{\prime}$, short-range $d_{xy}$-wave correlations dominate robustly across all fillings. In contrast, at larger $t^{\prime}$, the dominant pairing symmetry evolves with filling—from third-nearest-neighbor $d_{x^2 - y^2}^{(2)}$-wave at low $n$ to nearest-neighbor $d_{x^2 - y^2}^{(1)}$-wave at high $n$.

This paper is organized as follows: Section~\ref{Method and Model} introduces the model Hamiltonian and numerical methodology. In Section~\ref{Phase diagram}, we present the global $n$–$\alpha$ phase diagram, distinguishing CPBM from competing phases via momentum- and real-space diagnostics. Section~\ref{Fermi Surface} explores the geometric rotation and coupling of Fermi and BS. Section~\ref{Channels} identifies the dominant pairing symmetry within the CPBM phase.
In Sec.\ \ref{Discussion and Conclusion}, we give the discussion and conclusion.

\section{Method and Model}
\label{Method and Model}
We study the attractive Hubbard model on a two-dimensional square lattice with spin-dependent NN and NNN hopping, given by
\begin{equation}
H = -\sum_{\langle i,j \rangle, \sigma} t_{ij,\sigma} \, c_{i\sigma}^\dagger c_{j\sigma} 
    - \sum_{\langle\langle i,j \rangle\rangle, \sigma} t^{\prime}_{ij,\sigma} \, c_{i\sigma}^\dagger c_{j\sigma}
    + U \sum_i n_{i\uparrow} n_{i\downarrow}.
\end{equation}
Here, $c_{i\sigma}^\dagger$ creates a fermion with spin $\sigma$ at site $i$, and $n_{i\sigma} = c_{i\sigma}^\dagger c_{i\sigma}$. The spin-dependent hopping amplitudes are defined as $t_{x,\uparrow} = t$, $t_{y,\uparrow} = \alpha t$, and similarly for the NNN direction, $t^{\prime}_{x+y,\uparrow} = t^{\prime}$, $t^{\prime}_{x-y,\uparrow} = \alpha t^{\prime}$. The down-spin amplitudes follow by spin inversion symmetry: $t_{x,\downarrow} = \alpha t$, $t_{y,\downarrow} = t$, $t^{\prime}_{x+y,\downarrow} = \alpha t^{\prime}$, and $t^{\prime}_{x-y,\downarrow} = t^{\prime}$. For convenience,we set the NN hopping $t$ = 1. The dimensionless parameter $\alpha \in [0,1]$ encodes the degree of spin anisotropy; the model reduces to the isotropic case for $\alpha = 1$, and becomes maximally anisotropic for $\alpha = 0$.

Our study systematically explores the full range of anisotropy and electron filling, sweeping $\alpha \in [0,1]$ and $n \in [0,1]$ to map out the evolution of the CPBM phase and its associated BS. Unless otherwise specified, the interaction strength is fixed at $U = -3$, which lies in the intermediate regime of the BCS–BEC crossover and is known to stabilize non-superfluid pair phases without inducing severe sign problems in numerical simulations. We primarily focus on the moderate NNN hopping ratio $t^{\prime}/t = 0.2$~\cite{Cao2025}, motivated by its relevance to cuprate band structures~\cite{Huang2001} and the relatively unexplored role of $t^{\prime}$ at fixed anisotropy across fillings. Additional simulations at $t^{\prime}/t = 0.5$ and $0.8$ serve as comparative benchmarks to clarify the effect of stronger kinetic frustration.

Ground-state properties are computed using the CPQMC method, which systematically approaches the ground-state wavefunction through imaginary-time evolution of the many-body system, employing a trial-wavefunction-based constraint to mitigate the fermion sign problem via the constrained path approximation~\cite{Zhang1997,PhysRevLett.74.3652,PhysRevB.55.7464}. Unless otherwise noted, all presented data are obtained on a $16 \times 16$ lattice, which is adequate for capturing the characteristics of BS structures.

As a key observable, we evaluate the effective momentum-resolved pairing distribution:
\begin{equation}
\small
N^{\text{eff}}_{s\text{-pair}}(\mathbf{k}) = \frac{1}{N} \sum_{i,j} e^{i \mathbf{k} \cdot (\mathbf{r}_i - \mathbf{r}_j)} \left[ \langle \Delta_s^\dagger(i) \Delta_s(j) \rangle - \langle c_{i\uparrow}^\dagger c_{j\uparrow} \rangle \langle c_{i\downarrow}^\dagger c_{j\downarrow} \rangle \right],
\end{equation}
where $\Delta_s^\dagger(i) = c_{i\uparrow}^\dagger c_{i\downarrow}^\dagger$ creates an on-site $s$-wave singlet. This quantity captures the momentum structure of connected pairing correlations. Condensation of $N^{\text{eff}}_{s\text{-pair}}(\mathbf{k})$ at nonzero momentum signals the formation of a Bose surface (BS)—a hallmark of the CPBM phase. In our analysis, we use the distribution’s peak position, shape, and symmetry to trace the evolution of the BS across parameter space.We also define the real-space correlation of the effective on-site $s$-wave pairing as $C^{\rm eff}_{\mathrm s-pair}(i,j) = \langle {\Delta}_{s}^{\dagger}(i) {\Delta}_{s}(j) \rangle - G^{\uparrow}_{i,j}G^{\downarrow}_{i,j}$, where $G^{\sigma}_{i,j}= \langle c_{i\sigma}c^\dagger_{j\sigma} \rangle$ denotes the single-particle Green’s function, and the second term removes uncorrelated contributions to isolate genuine pairing effects.

To study the density correlations, we defined the charge structure factor in particle-hole channel, 
\begin{equation}
N_{\mathrm c}({\bf k}) = (1/N)\sum_{i,j} \mbox{exp}[i{\bf k}({\bf r}_i-{\bf r}_j)]\, \langle {n}_i {n}_j \rangle,
\label{nskpair}
\end{equation}
Here, the density number operator is defined as
${n}_i = \sum_{\sigma} c^\dagger_{i \sigma} c_{i \sigma}$.The correlation function of charge density wave in real space are defined as $C_{\rm CDW} = \langle {n}_i {n}_j \rangle$.

To characterize the internal structure of CPBM phase, we evaluate the effective two-boson momentum distribution $P^{\text{eff}}_\zeta(\mathbf{k})$, which serves as a diagnostic for symmetry-resolved pair-pair correlations. Following Ref.~\cite{Cao2025}, we define this correlator based on connected four-fermion observables built from NN bosonic pairs $b_i^\dagger = c_{i\uparrow}^\dagger c_{i\downarrow}^\dagger$, with
\begin{equation}
P^{\text{eff}}_\zeta(\mathbf{k}) = \frac{1}{N} \sum_{i,j} e^{i \mathbf{k} \cdot (\mathbf{r}_i - \mathbf{r}_j)} C^{\text{eff}}_\zeta(i,j),
\end{equation}
where the real-space correlator $C^{\text{eff}}_\zeta(i,j)$ is defined over displaced bonds labeled by symmetry channel $\zeta$:
\begin{align}
C^{\text{eff}}_\zeta(i,j) = 
\sum_{\delta_\zeta, \delta'_\zeta} 
\Big[ 
& \langle b_i^\dagger b_{i+\delta_\zeta}^\dagger 
         b_j b_{j+\delta'_\zeta} \rangle \nonumber \\
& - G^\uparrow_{i,j} G^\uparrow_{i+\delta_\zeta, j+\delta'_\zeta} 
    G^\downarrow_{i,j} G^\downarrow_{i+\delta_\zeta, j+\delta'_\zeta} 
\Big].
\end{align}

The choice of displacement vectors $\delta_\zeta$ specifies the internal pairing symmetry: for $s$-wave and $d_{x^2 - y^2}^{(1)}$-wave, $\delta_\zeta$ spans NN sites; for $d_{xy}$-wave, it spans NNN sites; and for $d_{x^2 - y^2}^{(2)}$-wave, it spans third-nearest-neighbor bonds. These channels correspond to local, NN, and extended $d$-wave pairing patterns and are used to extract symmetry-resolved pairing strength as a function of filling.
In our calculations, we focus on the peak value $P^{\text{eff}}_\zeta(\mathbf{k}_{\text{max}})$ for each channel and monitor its evolution with filling. 

\section{Result}
\label{Result}
\subsection{Phase diagram of CPBM with next-nearest-neighbor hopping}
\label{Phase diagram}
\begin{figure}[tb!]
    \centering
    \includegraphics[width=\linewidth]{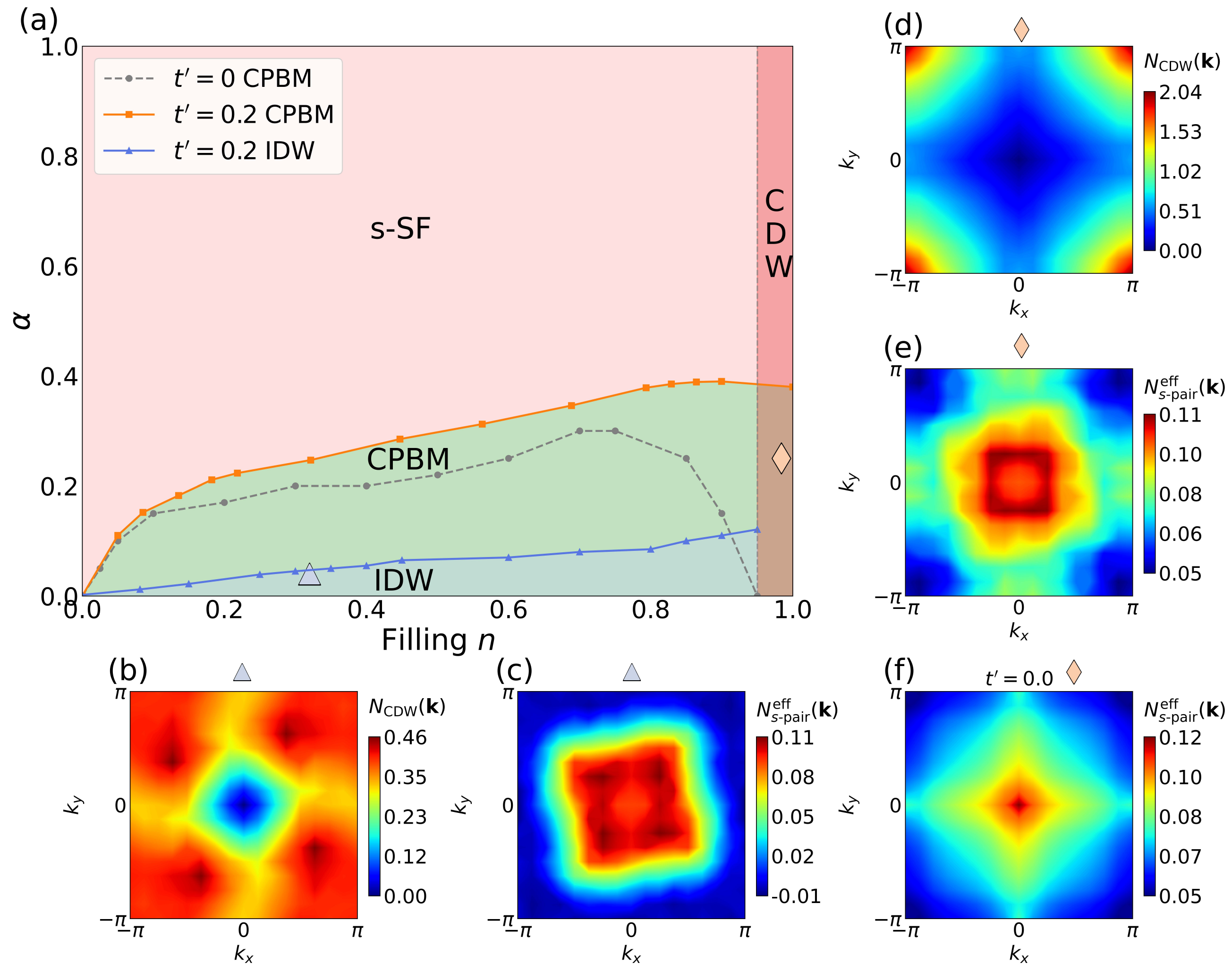} 
    \caption{(Color online) Phase diagram. (a) Zero-temperature phase diagram in the $n$--$\alpha$ plane at $t^{\prime}/t = 0.2$. The CPBM–s-SF boundary is shown by the orange solid line with square markers, and the corresponding $t^{\prime} = 0$ boundary by the gray dashed line with circles, both determined from $N^{\text{eff}}_{s\text{-pair}}(\mathbf{k})$. The IDW region, marked by the blue solid line with triangles, is identified via peaks in $N_c(\mathbf{k})$ near $(2k_F, 2k_F)$ and terminates at $n \simeq 0.95$. The commensurate CDW phase appears beyond this point, with a dominant peak at $(\pi, \pi)$, and can coexist with CPBM only for $t^{\prime}/t = 0.2$. The diamond and triangle symbols denote the specific parameter sets illustrated in panels (b), (c), (d), (e) and (f).
    (b) $N_c(\mathbf{k})$ at $(n, \alpha) = (0.320, 0.05)$ showing IDW peaks near $(2k_F, 2k_F)$.  
    (c) $N^{\text{eff}}_{s\text{-pair}}(\mathbf{k})$ at the same point, exhibiting a finite-momentum BS characteristic of CPBM phase.  
    (d) $N_c(\mathbf{k})$ at $(n, \alpha) = (0.982, 0.20)$ showing a sharp commensurate peak at $(\pi, \pi)$, consistent with CDW.  
    (e), (f) $N^{\text{eff}}_{s\text{-pair}}(\mathbf{k})$ at $(n, \alpha) = (0.982, 0.20)$ for $t^{\prime}/t = 0.2$ and $t^{\prime} = 0$, respectively. CPBM persists at high filling for finite $t^{\prime}$, but collapses when $t^{\prime} = 0$, underscoring the essential role of NNN hopping in stabilizing the Bose metallic phase. }
    \label{fig1}
\end{figure}

To establish the global phase structure, we map out the zero-temperature phase diagram as a function of filling $n$ and spin anisotropy $\alpha$ at $t^{\prime}/t = 0.2$ using CPQMC simulations [Fig.~\ref{fig1}(a)]. Four distinct phases are identified based on the momentum-space structure of the charge density and pairing observables: s-SF, CPBM, incommensurate density wave (IDW), and CDW. These phases exhibit distinct distribution patterns in the phase diagram: The $s$-SF phase occupies all regions not otherwise identified as CPBM, IDW, or CDW. The s-SF predominantly emerges at weak anisotropy and across a wide filling range, characterized by zero-momentum condensation where the effective NN $s$-wave pairing distribution function $N_{\text{s-pair}}^{\text{eff}}(\mathbf{k})$ exhibits a sharp condensed peak at $\mathbf{k} = (0, 0)$.

As the system enters the strong anisotropy regime, the CPBM phase emerges, characterized by a BS located at finite momentum in $N_{s\text{-pair}}^{\text{eff}}(\mathbf{k})$. For $t^{\prime}/t = 0.2$, the CPBM is particularly robust at high filling, reaching its maximum extent near $n \approx 0.90$ and persisting up to $n = 1.0$ for $\alpha \lesssim 0.38$, as delineated by the orange solid line in Fig.~\ref{fig1}(a). In contrast, the gray dashed line reproduces the corresponding CPBM–s-SF boundary at $t^{\prime} = 0$~\cite{Cao2024}, where the phase terminates near $n \sim 0.95$. This comparison indicates that finite NNN hopping $t^{\prime}$ significantly enhances the stability of the Bose-metal phase in the high-density regime by frustrating FS nesting and suppressing competing charge ordering [Fig.~\ref{fig3}(e)].
At low fillings and strong anisotropy, the system enters the IDW phase, identified by a peak in $N_c(\mathbf{k})$ near $\mathbf{Q} \approx (2k_F, 2k_F)$. As $n$ increases, this incommensurate peak gradually shifts toward the commensurate wavevector $\mathbf{Q} = (\pi, \pi)$, signaling a smooth evolution into the CDW phase near $n \sim 0.95$. The blue solid line in Fig.~\ref{fig1}(a) marks the boundary of the IDW regime, which terminates at this transition. For $n \gtrsim 0.95$, the system develops robust CDW order, as evidenced by the sharp peak in $N_c(\mathbf{k})$ at $(\pi, \pi)$ [Fig.~\ref{fig1}(d)].

Representative momentum-resolved observables confirm this phase structure. At $(n, \alpha) = (0.320, 0.05)$, $N_c(\mathbf{k})$ shows IDW peaks [Fig.~\ref{fig1}(b)], while $N_{s\text{-pair}}^{\text{eff}}(\mathbf{k})$ reveals finite-momentum condensation [Fig.~\ref{fig1}(c)], confirming coexistence of CPBM and density-wave correlations. At $(n, \alpha) = (0.982, 0.20)$, $N_c(\mathbf{k})$ is sharply peaked at $(\pi, \pi)$, indicative of commensurate CDW order [Fig.~\ref{fig1}(d)]. In particular, Figs.~\ref{fig1}(e) and \ref{fig1}(f) compare $N_{s\text{-pair}}^{\text{eff}}(\mathbf{k})$ at $(n, \alpha) = (0.982, 0.20)$ for $t^{\prime}/t = 0.2$ and $t^{\prime} = 0$. The CPBM phase remains robust in the presence of finite $t^{\prime}$ [Fig.~\ref{fig1}(e)], but disappears entirely at $t^{\prime} = 0$ [Fig.~\ref{fig1}(f)], underscoring the critical role of kinetic frustration from $t^{\prime}$ in breaking nesting and stabilizing the Bose metal against competing orders.

Building on the global phase diagram, we now turn to microscopic criteria that distinguish the CPBM from the s-SF based on momentum and real-space pairing structure. To systematically identify the CPBM phase and distinguish it from the s-SF, we analyze the momentum-space pairing structure and real-space decay of $N^{\text{eff}}_{s\text{-pair}}(r)$.
\begin{figure}[tb!]
    \includegraphics[width=\linewidth]{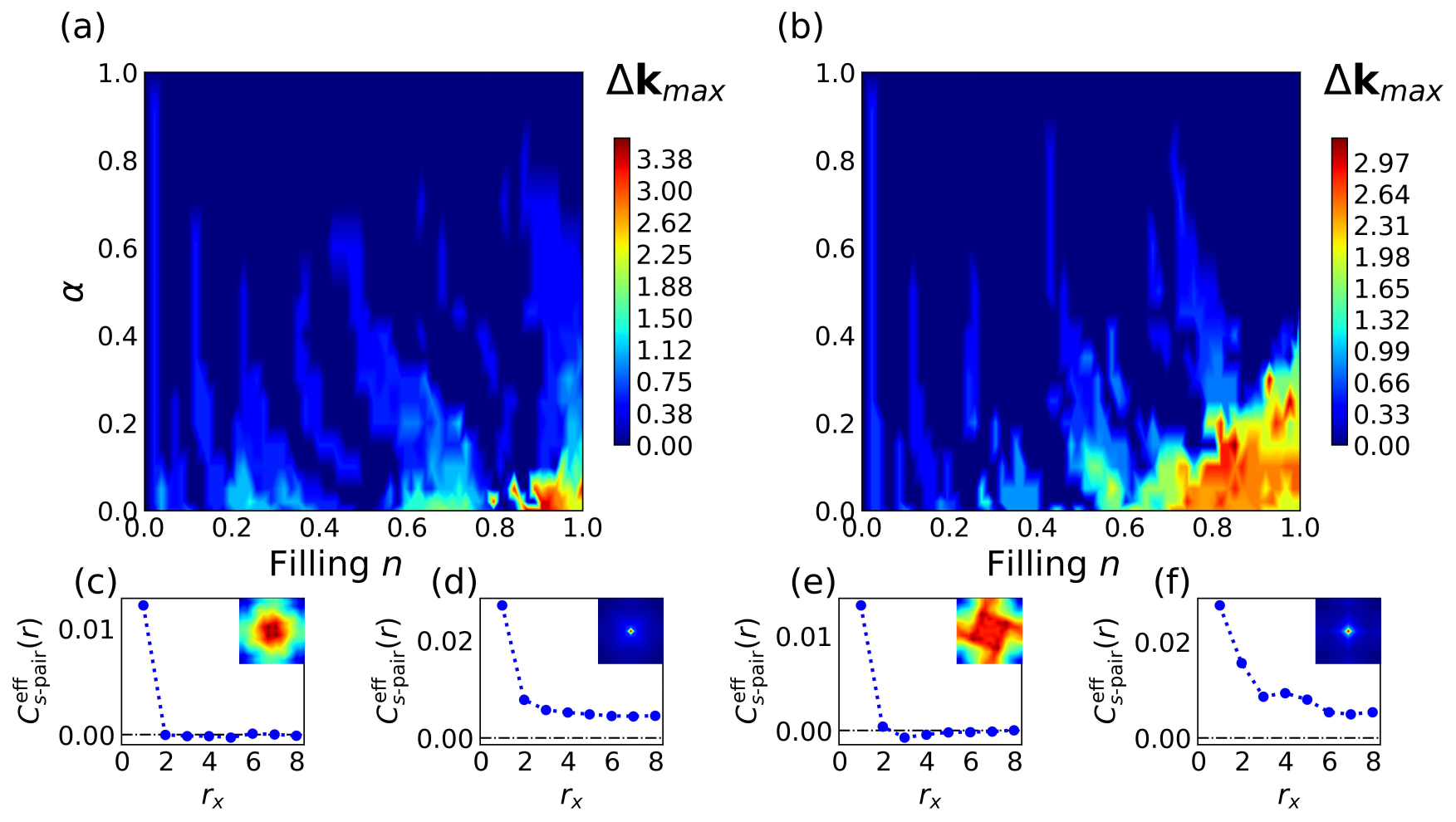}
    \caption{(Color online) 
    Characterization of the CPBM and $s$-SF phases via momentum-resolved and real-space pairing observables. (a),(b) Colormaps of $\Delta k_{\text{max}} = \sqrt{k_{\text{max}}^2(x) + k_{\text{max}}^2(y)}$, extracted from $N^{\text{eff}}_{s\text{-pair}}(\mathbf{k})$, as a function of $n$ and $\alpha$ for $t^{\prime}/t = 0.2$ and $0.8$, respectively. Blue regions indicate $s$-wave pairing with condensation at $\mathbf{k} = 0$, while finite values reflect the presence of a BS characteristic of CPBM. As $t^{\prime}$ increases, CPBM regions (red/yellow) shift toward higher filling, consistent with enhanced frustration. (c), (d) Real-space decay of $N^{\text{eff}}_{s\text{-pair}}(r_x)$ for $t^{\prime}/t = 0.2$. At $(n, \alpha) = (0.633, 1.00)$ (c), the correlations persist over long distances, confirming $s$-SF behavior; inset shows a sharp peak at $\mathbf{k} = 0$. At $(0.938, 0.25)$ (d), the decay is rapid and the inset shows finite-momentum condensation, identifying a CPBM phase. (e), (f) Same analysis for $t^{\prime}/t = 0.8$. Panel (e), at $(0.938, 0.90)$, exhibits short-range pairing and a square-like BS. Panel (f), at $(0.914, 0.20)$, displays long-range coherence and a zero-momentum condensate, consistent with $s$-SF. }
    \label{fig2}
\end{figure}
Figs.~\ref{fig2}(a) and \ref{fig2}(b) present colormaps of the peak position $\Delta k_{\text{max}}$ of $N^{\text{eff}}_{s\text{-pair}}(\mathbf{k})$ across the $(n, \alpha)$ plane for $t^{\prime}/t = 0.2$ and $0.8$, respectively. A value $\Delta k_{\text{max}} = 0$ (blue regions) indicates condensation at $\mathbf{Q} = (0,0)$ and thus conventional s-SF, while $\Delta k_{\text{max}} \neq 0$ (colored regions) reflects finite-momentum condensation characteristic of CPBM. As $t^{\prime}$ increases, the red/yellow regions become more prominent and concentrated at high fillings, signaling enhanced CPBM stability. This trend results from the geometric frustration introduced by NNN hopping, which suppresses FS nesting and enhances CPBM stability.

To validate the nature of these regions, we further examine representative real-space pairing correlations. Fig.~\ref{fig2}(c) shows $C^{\text{eff}}_{s\text{-pair}}(r_x)$ at $(n, \alpha) = (0.633, 1.00)$, deep in the s-SF phase, where the correlation decays slowly and remains finite at large $r_x$; the inset confirms that $N^{\text{eff}}_{s\text{-pair}}(\mathbf{k})$ is sharply peaked at $\mathbf{k} = 0$. In contrast, panel (d) at $(n, \alpha) = (0.938, 0.25)$ shows rapid decay of $N^{\text{eff}}_{s\text{-pair}}(r_x)$ and a momentum distribution peaked at finite $\mathbf{k}$ values, consistent with short-range CPBM behavior. Similar diagnostics are presented for $t^{\prime}/t = 0.8$ in Fig.~\ref{fig2}(e) and \ref{fig2}(f). At $(n, \alpha) = (0.938, 0.20)$, the pairing remains short-ranged with a square-like BS, again confirming CPBM, while at $(n, \alpha) = (0.914, 0.90)$, the reappearance of a long-range tail in $C^{\text{eff}}_{s\text{-pair}}(r_x)$ and a momentum-space distribution peaked at $\mathbf{k} = 0$ signal a recovery of conventional s-wave superfluid order.

Taken together, these observations demonstrate that finite-momentum pairing (nonzero $\Delta k_{\text{max}}$) and short-range real-space decay unambiguously identify the CPBM phase. Despite finite-size fluctuations on the $16 \times 16$ lattice, we expect the CPBM regions to become contiguous in the thermodynamic limit ($N \rightarrow \infty$)~\cite{Cao2024}. The cooperative use of momentum- and real-space observables thus provides a robust criterion for CPBM detection across the full parameter space.

While the earlier analysis focused on pairing correlations to distinguish CPBM from s-SF, we now turn to charge-ordering phenomena to characterize the transition between IDW and CDW density waves. To this end, we analyze the momentum dependence of the charge structure factor $N_c(\mathbf{k})$ across parameter space.

Fig.~\ref{fig3}(a) shows the evolution of $\Delta k_{\text{max}} = \sqrt{k_{\text{max}}^2(x) + k_{\text{max}}^2(y)}$, which characterizes the position of the dominant peak in momentum space. When $\Delta k_{\text{max}} \approx \sqrt{2}\pi$, the peak resides at $\mathbf{Q} = (\pi,\pi)$, signaling the formation of commensurate CDW order. Deviations from this value, especially continuous shifts with filling, indicate finite-wavevector ordering consistent with IDW behavior. To further quantify the peak sharpness, we define the ratio $R_p = (p - \bar{p}) / p$, where $p$ is the maximum value of $N_c(\mathbf{k})$ and $\bar{p}$ is the average over neighboring momentum values. As shown in Fig.~\ref{fig3}(b), $R_p$ becomes large in regions with IDW, particularly for $0.2 \lesssim n \lesssim 0.9$ under strong anisotropy. Interestingly, for $n > 0.95$, $R_p$ again increases in a narrow band near $\alpha \sim 0.1$–$0.4$, where $\Delta k_{\text{max}}$ approaches $(\pi,\pi)$, indicating the reemergence of commensurate CDW order in the high-filling limit.

These global trends are corroborated by real-space data. Fig.~\ref{fig3}(c) displays the charge–charge correlation function $C_{\text{CDW}}(r_x)$ at $(n, \alpha) = (0.320, 0.05)$, showing long-range oscillations with non-integer periodicity. The inset confirms an off-commensurate peak in $N_c(\mathbf{k})$ near $\mathbf{Q} \approx (2k_F, 2k_F)$, consistent with IDW behavior. In contrast, Fig.~\ref{fig3}(d) at $(n, \alpha) = (0.953, 0.15)$ exhibits regular charge modulation with period $2a$, and a sharp peak pinned at $(\pi,\pi)$ in momentum space, characteristic of a CDW state~\cite{Cao2024}. Notably, CDW correlations are a central feature of the cuprates~\cite{Comin2016} and other families of unconventional superconductors~\cite{Teng2022}, making their emergence in the CPBM context particularly relevant
\begin{figure}[tb]
    \includegraphics[width=\linewidth]{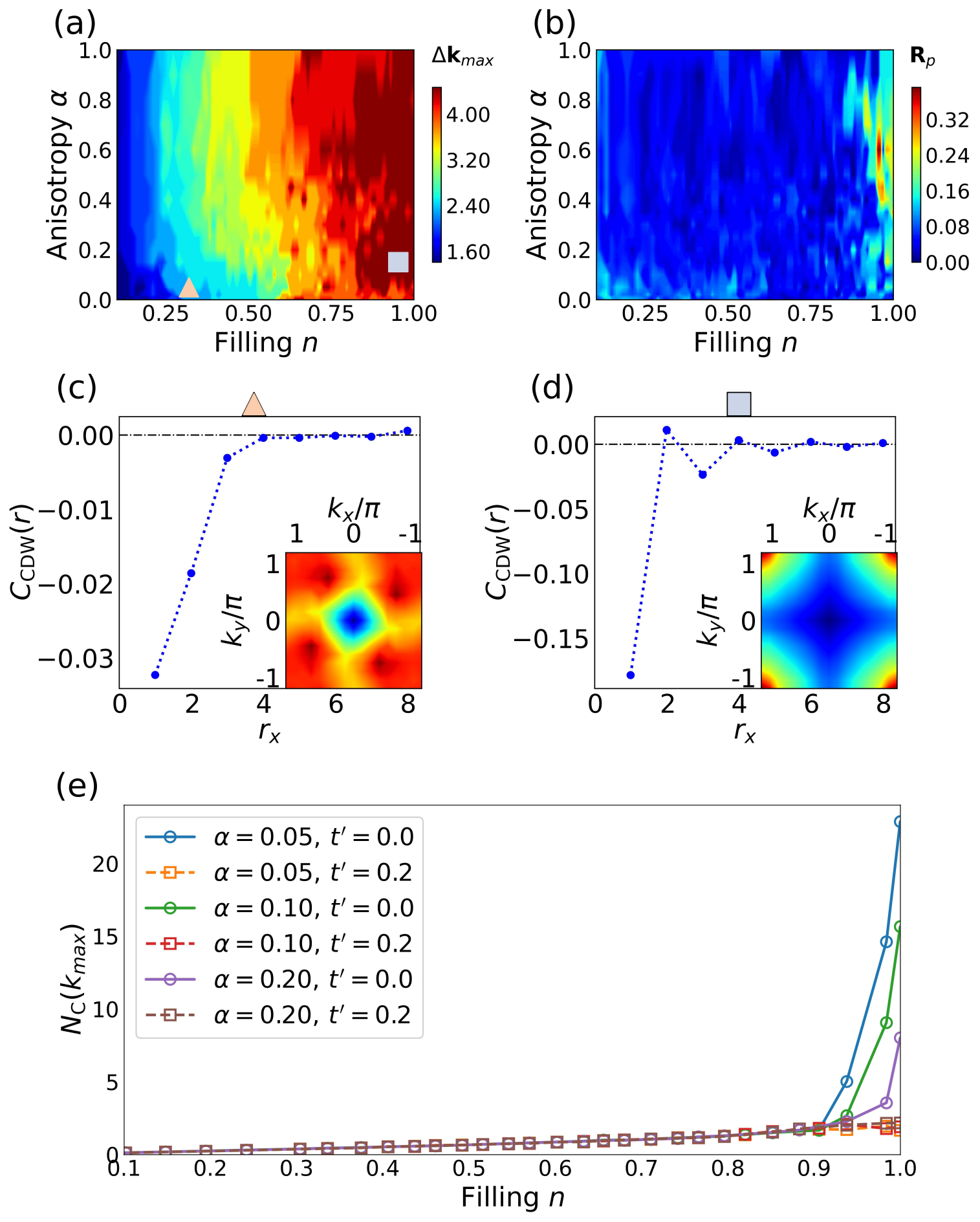}
    \caption{(Color online) Identification of IDW and CDW phases from charge observables. (a) Colormap of $\Delta k_{\text{max}}$, the peak position of $N_c(\mathbf{k})$, across $(n,\alpha)$. Values near $\sqrt{2}\pi$ (red) indicate CDW order at $(\pi,\pi)$.The square and triangle symbols denote the specific parameter sets illustrated in panels  (c) and (d).(b) Sharpness ratio $R_p = (p - \bar{p}) / p$, highlighting well-defined IDW/CDW peaks (large $R_p$) versus diffuse charge fluctuations ($R_p \approx 0$). (c), (d) Real-space correlator $C_{\text{CDW}}(r_x)$ for IDW [panel (c), $(0.320, 0.05)$] and CDW [panel (d), $(0.953, 0.15)$]. Insets: $N_c(\mathbf{k})$ peaks at $(2k_F,2k_F)$ and $(\pi,\pi)$, respectively.(e) Comparison of $N_c(\mathbf{k}_{\text{max}})$ vs.~$n$ for various $t^{\prime}$ and $\alpha$. For $n \lesssim 0.95$, all curves coincide, showing $t^{\prime}$ does not affect IDW. For $n > 0.95$, CDW grows sharply for $t^{\prime}=0$, but is suppressed for $t^{\prime}/t = 0.2$, consistent with CPBM–CDW competition at high filling.}
    \label{fig3}
\end{figure}
To further examine the effect of NNN hopping, Fig.~\ref{fig3}(e) compares the maximum CDW structure factor $N_c(\mathbf{k}_{\text{max}})$ versus filling $n$ for several combinations of anisotropy $\alpha$ and hopping strength $t^{\prime}$. For $n \lesssim 0.95$, the values for $t^{\prime} = 0$ and $t^{\prime} = 0.2$ remain nearly indistinguishable, indicating that moderate NNN hopping does not significantly alter IDW strength. However, for $n > 0.95$, the $t^{\prime} = 0$ cases exhibit a sharp increase in CDW amplitude, while the $t^{\prime} = 0.2$ cases remain nearly flat or slightly suppressed. This behavior suggests that in the high-filling regime, the CPBM phase stabilized by $t^{\prime} = 0.2$ acts to suppress CDW ordering.

Despite this suppression, panels Figs.~\ref{fig3}(b) and ~\ref{fig3}(d) indicate that CDW correlations still persist at reduced intensity in the presence of CPBM, implying possible coexistence or competition. These results confirm that the evolution of charge order with filling is sensitive to both anisotropy and kinetic frustration, and that the CPBM phase may act to frustrate long-range density-wave ordering in a filling-dependent manner.

Taken together, our analyses in both momentum and real space provide a consistent characterization of the charge-order evolution throughout the phase diagram. The IDW regime manifests as smoothly shifting incommensurate peaks and non-integer periodicity in $C_{\text{CDW}}(r_x)$, while the CDW phase locks into $(\pi, \pi)$ order with sharply peaked $N_c(\mathbf{k})$ and long-range correlations. Importantly, moderate NNN hopping not only suppresses CDW tendencies at high filling, but also significantly expands the CPBM regime toward both lower anisotropy and higher density. This enhanced stability allows the CPBM to emerge as a robust competing ground state, bridging regions that would otherwise be dominated by conventional superfluid or charge-ordered phases.

We perform complementary finite-size scaling analyses, as detailed in Appendix~\ref{appendix:S1}. Results from $L=8$–$16$ lattices exhibit consistent scaling behavior in both pairing and charge observables, suggesting that the key features of all four phases—CPBM, s-SF, IDW, and CDW—persist in the thermodynamic limit. These results support the extracted phase boundaries and lend further credence to the distinction between the CPBM phase and the symmetry-broken ordered states identified in this study. 

\subsection{Influence of $t^{\prime}$ on Fermi Surface}
\label{Fermi Surface}
Beyond its influence on charge-order competition, the NNN hopping $t^{\prime}$ also plays a fundamental role in modifying the underlying band geometry. To gain further insight into the role of $t^{\prime}$ in reshaping the fermionic landscape, we examine its impact on the Fermi-surface topology.

\begin{figure}[tb!]
    \includegraphics[width=1\linewidth]{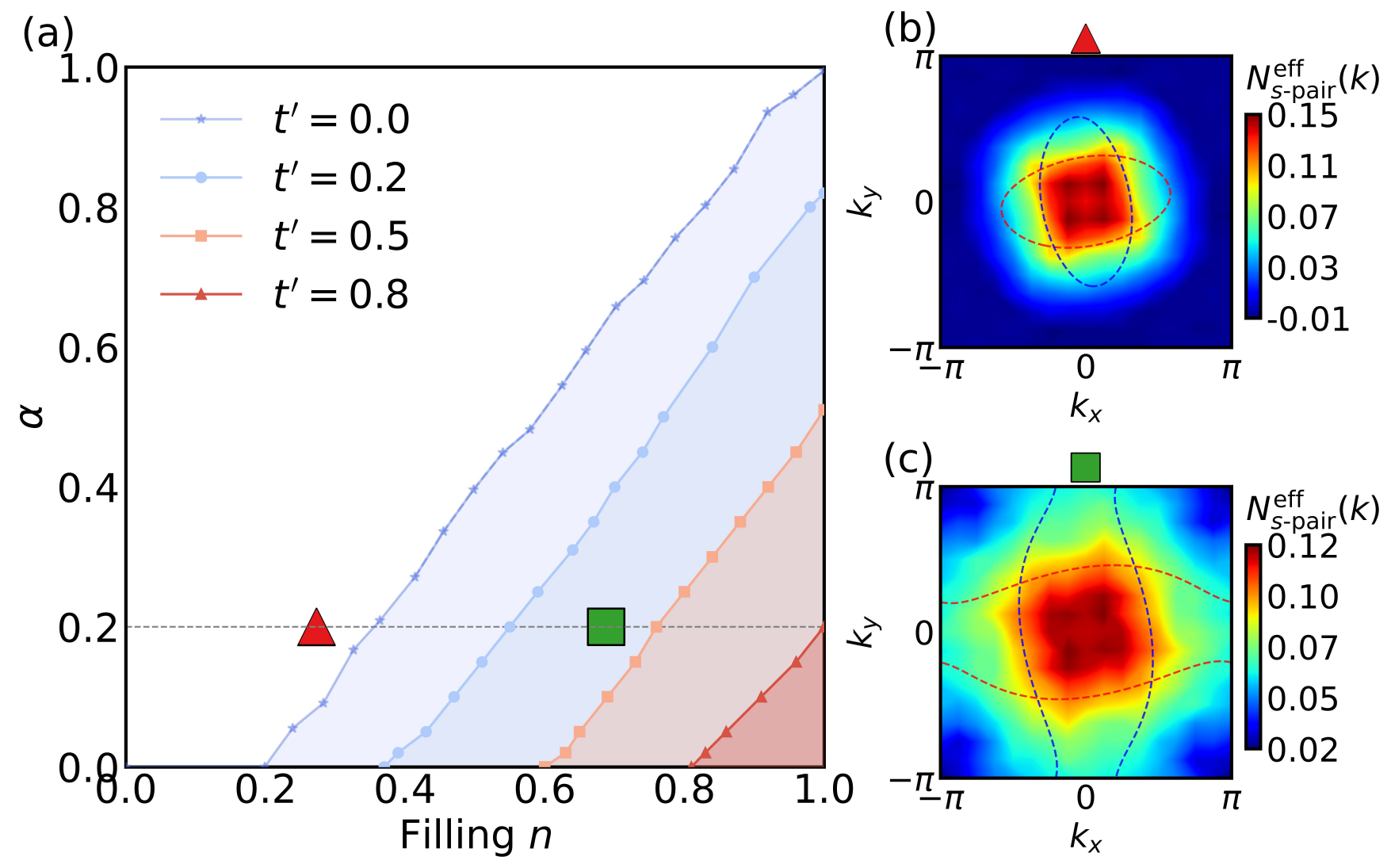}
    \caption{(Color online) Topology of the non-interacting FS and the
    corresponding effective onsite $s$-wave pair distribution. 
    (a) Boundary lines that separate closed (left side) and open (right side)
    FS in the $n$–$\alpha$ plane for
    $t^{\prime} = 0$ (blue-violet stars), $0.2$ (light-blue circles),
    $0.5$ (light-orange squares), and $0.8$ (red triangles).
    Shaded areas highlight the parameter region where the surface remains closed.
    The horizontal gray dashed line marks the fixed anisotropy $\alpha = 0.20$
    used in panels (b) and (c);
    the red triangle and green square indicates the specific parameter set shown in (b) [(c)]. 
    (b) Contour map of $N^{\text{eff}}_{s\text{-pair}}(\mathbf{k})$
    at $(t^{\prime},n,\alpha) = (0.2,\,0.273,\,0.20)$ together with the associated
    closed FS (red line for spin-$\uparrow$, blue line for spin-$\downarrow$). 
    (c) Same as (b) but for $(t^{\prime},n,\alpha) = (0.2,\,0.688,\,0.20)$,
    where the FS is open.}
    \label{fig4}
\end{figure}

Fig.~\ref{fig4}(a) shows the critical lines $\alpha_c(n)$ that separate closed (left of each curve) from open (right of each curve) FS, obtained from the non-interacting dispersion for four representative values of $t^{\prime}$. As $t^{\prime}$ increases, the open–closed boundary shifts systematically toward larger filling, indicating that a higher carrier concentration is required to drive the surface open. For example, at fixed spin anisotropy $\alpha = 0.20$, the critical filling rises from $n_c \simeq 0.32$ for $t^{\prime} = 0$ to $n_c \simeq 0.55$ for $t^{\prime} = 0.2$, $n_c \simeq 0.76$ for $t^{\prime} = 0.5$, and $n_c \simeq 0.98$ for $t^{\prime} = 0.8$. Correspondingly, the portion of the $n$–$\alpha$ plane that supports an open FS shrinks steadily with increasing $t^{\prime}$.

To illustrate the microscopic consequences of this transition, Figs.~\ref{fig4}(b) and \ref{fig4}(c) inspect two points on opposite sides of the $t^{\prime} = 0.2$ boundary at $\alpha = 0.20$. For the lower filling $n = 0.273$ (red marker in Fig.~\ref{fig4}(a)), the FS is closed, and the ridge in $N^{\text{eff}}_{s\text{-pair}}(\mathbf{k})$ forms a square-like BS enclosing the Brillouin-zone center. Raising the filling to $n = 0.688$ (green marker) moves the system to the open side of the boundary; the ridge now extends along the open contour yet remains continuous, demonstrating that the Cooper-pair Bose-metal state survives the topological change. 

These results indicate that while $t^{\prime}$ determines where the FS switches from closed to open, the robust square-shaped enhancement in the pairing channel—and therefore the stability of the CPBM phase—is governed primarily by the spin-anisotropy parameter $\alpha$ and the filling $n$, rather than by whether the underlying fermionic contour is closed or open.

While the previous analysis focused on the open–closed topology of the FS, we now turn to its geometric deformation and quantify the associated rotation. Figs.~\ref{fig4}(b) and~\ref{fig4}(c) reveals pronounced distortion of both closed and open FS relative to the crystal axes, particularly with respect to the $k_y = 0$ line. To quantify this rotational behavior, we define a FS rotation angle $\theta_F$ using the following procedure: first, we identify the four intersection points between the spin-up and spin-down FS in momentum space. We then compute the midpoints of the two point pairs lying at $k_y > 0$ and $k_y < 0$, respectively, and finally determine the angle $\theta_F$ between the line connecting these midpoints and the $k_y = 0$ axis.

\begin{figure}[b!]
    \includegraphics[width=0.85\linewidth]{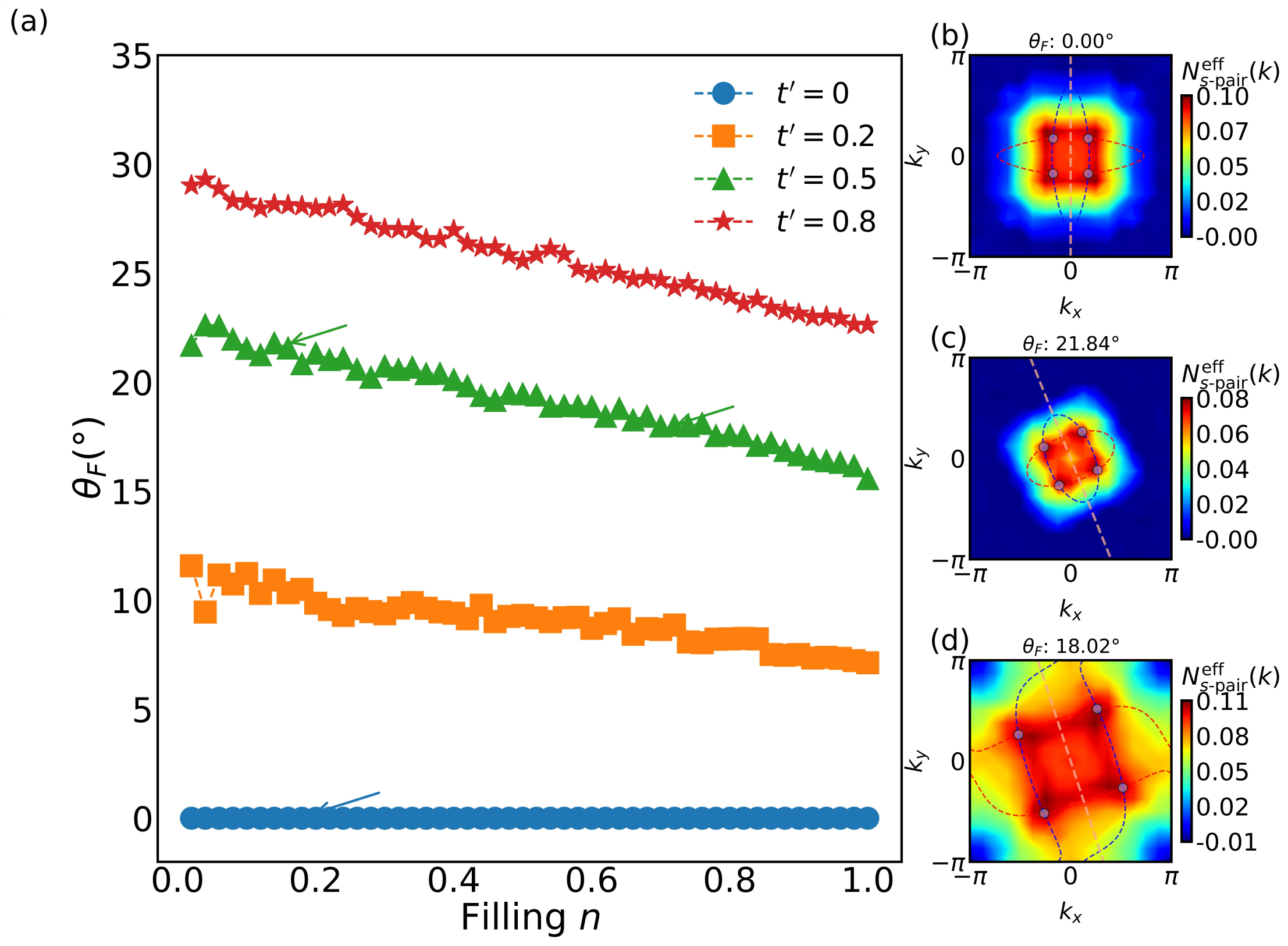}
    \caption{(Color online) Fermi-surface rotation as a function of NNN hopping and filling.
    (a) Rotation angle $\theta_F $ (degrees) versus filling $n$ at fixed anisotropy $\alpha=0.10$ for
    $t^{\prime}=0$ (blue circles), $0.2$ (orange squares), $0.5$ (green triangles), and $1.0$ (red pentagons).
    The three arrows point to the parameter sets illustrated in panels (b)–(d).
    (b)–(d) Contour plots of $N^{\text{eff}}_{s\text{-pair}}(\mathbf{k})$ together with the associated
    spin-resolved non-interacting FS.
    Panel (b) shows the unrotated case $(t^{\prime},n,\alpha)=(0,\,0.195,\,0.10)$ with $\theta_F=0^{\circ}$,
    while panels (c) and (d) display rotated contours at $(0.5,\,0.164,\,0.10)$
    with $\theta_F=21.84^{\circ}$ and $(0.5,\,0.719,\,0.10)$ with $\theta=18.02^{\circ}$, respectively.
    The angle $\theta_F$ is defined as the angle between the $k_y=0$ axis and the line joining
    the midpoint of the two spin-resolved intersection points at $k_y>0$
    to the midpoint of the corresponding pair at $k_y<0$.}
    \label{fig5}
\end{figure}

The systematic evolution of $\theta_F$ with electron filling is presented in Fig.~\ref{fig5}(a), where we fix the anisotropy at $\alpha = 0.10$. In the absence of NNN hopping ($t^{\prime} = 0$), the FS remains perfectly aligned with the lattice axes, i.e., $\theta_F = 0^\circ$, as illustrated in Fig.~\ref{fig5}(b) for $n = 0.195$. This behavior reflects the preserved $C_2$ rotational symmetry of the band structure, which contains no cross-coupling terms between orthogonal momentum directions. 
However, introducing finite NNN hopping ($t^{\prime} \ne 0$) qualitatively alters this picture. The hopping-induced terms $\cos(k_x \pm k_y)$ explicitly couple orthogonal momenta, breaking the rotational symmetry and inducing a global rotation of the FS. Two clear trends emerge. First, at any fixed $t^{\prime} \ne 0$, the rotation angle $\theta_F$ decreases monotonically as filling increases. For instance, with $t^{\prime} = 0.5$, $\theta_F$ reduces from $21.84^{\circ}$ at $n = 0.164$ [Fig.~\ref{fig5}(c)] to $18.02^{\circ}$ at $n = 0.719$ [Fig.~\ref{fig5}(d)]. This trend can be understood as a filling-induced outward expansion of the FS, which diminishes the influence of the anisotropic terms and tends to restore axial alignment. Second, for any fixed filling, the rotation angle $\theta_F$ increases monotonically with $t^{\prime}$, with larger NNN hopping amplitudes inducing stronger deviations from lattice-aligned FS; Specifically, $\theta_F$ exhibits a monotonic increase as $t^{\prime}/t$ increases from $0.2$ to $0.8$.

We further examine how $\theta_F$ varies with spin anisotropy $\alpha$ and find a clear density dependence. For low fillings ($n \lesssim 0.5$), the FS rotation angle remains largely insensitive to $\alpha$, suggesting that the band geometry is predominantly shaped by $t^{\prime}$ and the kinetic structure. In contrast, at higher fillings ($n = 0.800$ and $0.900$), $\theta_F$ exhibits a pronounced suppression with increasing $\alpha$—especially for $t^{\prime}/t = 0.5$ and $0.8$—indicating that strong spin anisotropy tends to restore axial alignment in the high-density regime by effectively reducing the impact of cross-hopping terms.

This rotational mechanism also helps explain key observations in Fig.~\ref{fig4}(a). As $t^{\prime}$ increases, the FS becomes more distorted in momentum space, tightening the condition for crossing the Brillouin zone boundary (e.g., at $k_x = \pi$ or $k_y = \pi$) and effectively delaying the formation of an open FS. Consequently, a higher carrier density is required to overcome the rotational offset and achieve an open topology. Moreover, this rotation displaces the nesting vectors from the ideal CDW wavevector $\mathbf{Q} = (\pi,\pi)$, thereby weakening the charge density wave susceptibility. The resulting geometric frustration of CDW correlations contributes to the enlarged CPBM phase region observed at moderate $t^{\prime}$, particularly $t^{\prime}/t = 0.2$, compared to the unfrustrated $t^{\prime} = 0$ case [Fig.~\ref{fig1}(a)]. In this way, NNN hopping not only enhances bosonic pairing but also suppresses its primary competitor—the CDW instability [Fig.~\ref{fig3}(e)]—thereby stabilizing the CPBM phase through a combined geometric and interaction-driven mechanism.

\begin{figure}[tb!]
    \includegraphics[width=\linewidth]{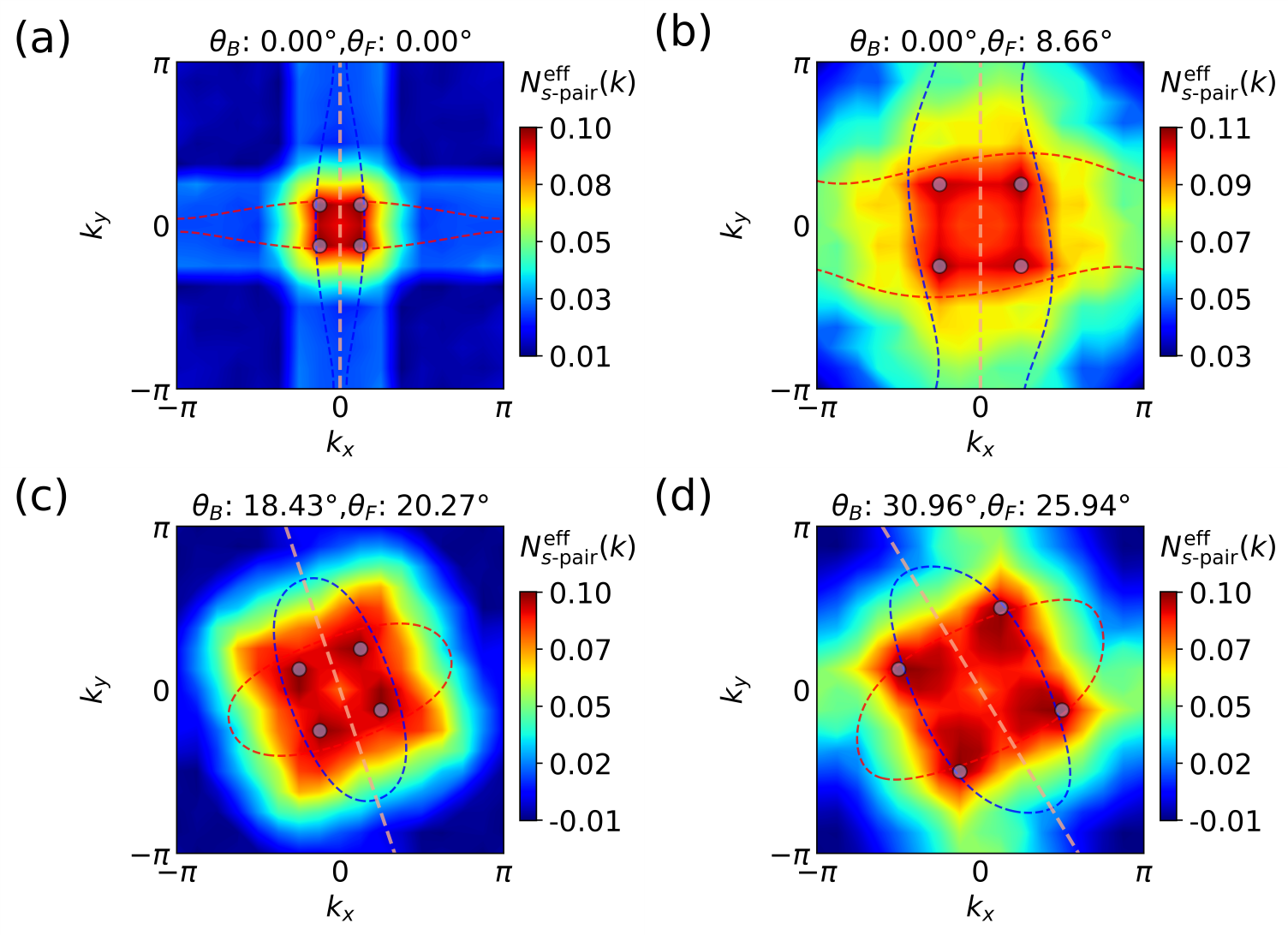}
    \caption{(Color online) The momentum-space structure of the BS \(N_{\text{s-pair}}^{\text{eff}}(\mathbf{k})\) and its correlation with the spin-dependent FS are shown. Each panel displays a color map of \(N_{\text{s-pair}}^{\text{eff}}(\mathbf{k})\) alongside the spin-up/spin-down Fermi contours (red and blue dashed lines). The BS rotation angle \(\theta_B\) is defined as the angle between the \(k_y = 0\) axis and the line connecting the midpoints of the upper and lower peak pairs of \(N_{\text{s-pair}}^{\text{eff}}(\mathbf{k})\) (analogous to the definition of \(\theta_F\) for the FS).
    (a) $t^{\prime}=0$,$n = 0.203$, $\alpha = 0.05$: no FS or BS rotation ($\theta_F = \theta_B = 0^\circ$).
    (b) $t^{\prime}/t = 0.2$, $n = 0.703$, $\alpha = 0.10$: $\theta_{\text{F}}\!\approx 8.66^{\circ}$, while the BS remains aligned along the lattice principal directions ($\theta_{\text{B}}\!\approx 0^{\circ}$).(c) $t^{\prime}/t = 0.5$, $n = 0.367$, $\alpha = 0.05$: $\theta_{\text{F}}\!\approx 18^{\circ}$ induces a smaller Bose-surface tilt $\theta_{\text{B}}\!\approx 8.4^{\circ}$.(d) $t^{\prime}/t = 0.8$, $n = 0.523$, $\alpha = 0.10$: at $\theta_{\text{FS}}\!\approx 26^{\circ}$ the BS co-rotates to $\theta_{\text{B}}\!\approx 31^{\circ}$, indicating a crossover where strong geometric frustration couples the two momentum-space structures.}
    \label{fig6}
\end{figure}

Having established the rotational behavior of the FS, we now investigate whether this deformation propagates to the structure of the BS in the CPBM phase. We now turn to the relationship between Fermi-surface rotation and the orientation of the BS. The BS in a CPBM state originates from momentum pairing between electrons on spin-split FS, with typical pair momentum $\mathbf{Q}_{\text{pair}} = \mathbf{k}_{F\uparrow} + \mathbf{k}_{F\downarrow}$~\cite{PhysRevB.78.054520}. This implies that Fermi-surface rotation may be inherited by the BS structure.

To quantify this behavior, we define a Bose-surface rotation angle $\theta_B$ in analogy to $\theta_F$. Specifically, we identify the four momentum-space maxima of $N^{\text{eff}}_{s\text{-pair}}(\mathbf{k})$, extract their midpoints at $k_y > 0$ and $k_y < 0$, and define $\theta_B$ as the angle between the line connecting these two midpoints and the $k_y = 0$ axis.

Fig.~\ref{fig6} presents this comparison across increasing values of $t^{\prime}$. When $t^{\prime} = 0$, both the Fermi and BS remain axis-aligned and unrotated ($\theta_F = \theta_B = 0^\circ$; Fig.~\ref{fig6}(a)). At small but finite $t^{\prime}$, the FS begins to rotate while the BS remains locked along the diagonals ($\theta_F = 8.66^\circ$, $\theta_B = 0^\circ$; Fig.~\ref{fig6}(b)). This indicates that mild FS distortion does not immediately affect the structure of the BS.However, when the FS rotation angle becomes sufficiently large, the BS begins to co-rotate with it [Figs.~\ref{fig6}(c) and~\ref{fig6}(d)]. For instance, at $t^{\prime}/t = 0.5$, the Fermi and BS rotate by $20.27^\circ$ and $18.43^\circ$, respectively; at $t^{\prime}/t = 0.8$, this co-rotation becomes more pronounced, with $\theta_F = 25.94^\circ$ and $\theta_B = 30.96^\circ$.

This threshold-like behavior suggests that $N^{\text{eff}}_{s\text{-pair}}(\mathbf{k})$ is initially governed by pair scattering between nearly orthogonal regions of the two spin-resolved FS. As the band distortion grows and the FS rotates further away from the crystal axes, the dominant pairing momentum—and hence the BS—gradually shifts in tandem.In other words, although the CPBM phase is not sensitive to whether the FS is open or closed, it can inherit the geometric rotation of the FS when the deformation becomes sufficiently pronounced~\cite{Cao2025}. This supports a microscopic picture in which finite-momentum Cooper pairing arises from the mismatch between spin-split Fermi contours, and their relative orientation is directly imprinted on the BS geometry.

Taken together, these results demonstrate that, when $t^{\prime} \ne 0$, the BS in the CPBM phase is not fixed by symmetry but instead evolves dynamically in response to the geometric rotation of the underlying FS. While the BS remains approximately locked along the crystal diagonals when the FS rotation angle is small, it gradually co-rotates with the FS as the latter becomes more distorted. This behavior supports a microscopic mechanism in which finite-momentum Cooper pairing emerges from anisotropic nesting between spin-split FS, whose mutual alignment is reflected in the BS structure. Thus, the geometry of the FS plays a dual role: not only suppressing CDW order, but also shaping the internal structure of the Bose-metal state.

\subsection{Evolution of Competing Bosonic Pairing Channels with Filling}
\label{Channels}
Having established the geometric features of the BS, we now investigate its internal structure by analyzing the symmetry of dominant bosonic pairing channels. To further elucidate the dominant pairing tendencies in the CPBM phase, we evaluate the peak values and real-space profiles of the momentum-resolved two-boson correlator $P^{\text{eff}}_\zeta(\mathbf{k})$ for four pairing symmetries: NN $s$-wave, $d_{xy}$-wave, NN $d_{x^2 - y^2}^{(1)}$-wave, and third-nearest-neighbor $d_{x^2 - y^2}^{(2)}$-wave. These channels correspond to distinct choices of bond vectors $\delta_\zeta$, as defined in Sec.~\ref{Method and Model}.

\begin{figure}[tb!]
    \centering
    \includegraphics[width=0.95\linewidth]{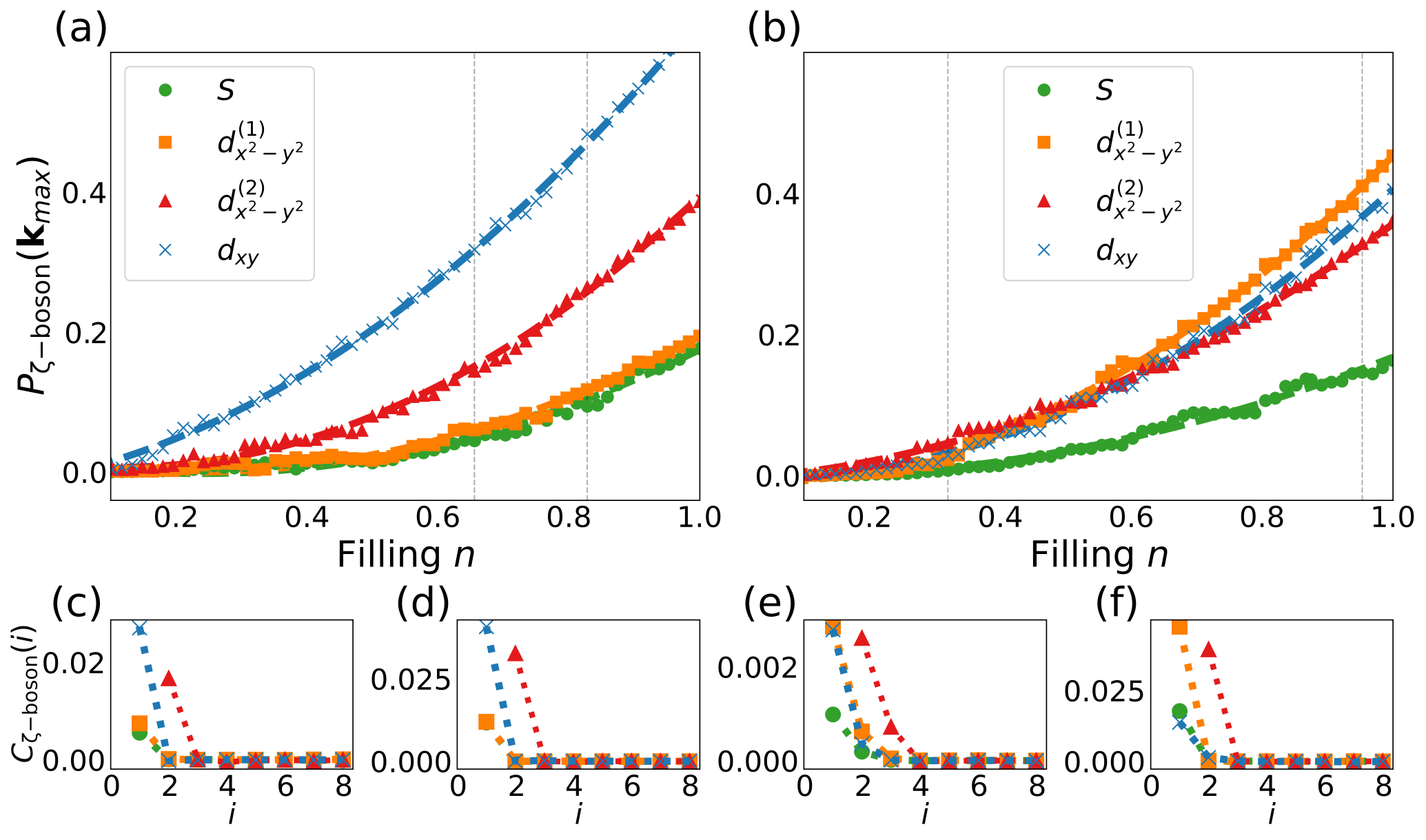}
       \caption{(Color online) Filling dependence and real-space profiles of bosonic pairing correlations in the CPBM phase at fixed anisotropy $\alpha = 0.05$. 
        (a) Peak value of $P^{\text{eff}}_\zeta(\mathbf{k}_{\text{max}})$ as a function of filling $n$ for $t^{\prime}/t = 0.2$. 
        The pairing channels include NN $s$-wave (green circles), NN $d_{x^2 - y^2}^{(1)}$-wave (orange squares), third-NN $d_{x^2 - y^2}^{(2)}$-wave (red triangles), and diagonal $d_{xy}$-wave (blue crosses). Vertical dashed lines indicate the fillings used in panels (c) and (d). 
        (b) Same as (a), but for $t^{\prime}/t = 0.8$, with vertical dashed lines marking the fillings used in panels (e) and (f). 
        (c), (d) Real-space correlations $C^{\text{eff}}_\zeta(i)$ for $t^{\prime}/t = 0.2$ at $n = 0.656$ and $n = 0.953$, respectively. 
        (e), (f) Corresponding results for $t^{\prime}/t = 0.8$ at $n = 0.648$ and $n = 0.953$. 
        All correlation functions decay rapidly with distance, confirming the short-range nature of bosonic pairing in the CPBM phase.}
    \label{fig7}
\end{figure}

Fig.~\ref{fig7}(a) shows the filling dependence of the peak value $P^{\text{eff}}_\zeta(\mathbf{k}_{\text{max}})$ for four symmetry channels at fixed anisotropy $\alpha = 0.05$ and moderate NNN hopping $t^{\prime}/t = 0.2$. Across the entire filling range, the $d_{xy}$-wave component remains dominant, followed by the third-nearest-neighbor $d_{x^2 - y^2}^{(2)}$-wave. The NN $d_{x^2 - y^2}^{(1)}$-wave component slightly exceeds the NN $s$-wave contribution, indicating that extended $d$-wave correlations dominate over $s$-wave pairing in this regime, classifying the system as a $d$-wave Bose metal\cite{PhysRevB.75.235116}.

Under stronger frustration ($t^{\prime}/t = 0.8$), the pairing structure undergoes a marked reorganization [Fig.~\ref{fig7}(b)]. At low fillings ($n \lesssim 0.55$), the $d_{x^2 - y^2}^{(2)}$-wave dominates. As $n$ increases, this component is overtaken first by $d_{x^2 - y^2}^{(1)}$-wave around $n \sim 0.55$, and subsequently by $d_{xy}$-wave, which becomes subdominant at high filling. Thus, for large $t^{\prime}$, the leading pairing symmetry depends sensitively on the particle filling, in contrast to the filling-independent $d_{xy}$-wave-dominant structure at smaller $t^{\prime}$.

\begin{figure}[tb!]
    \centering
    \includegraphics[width=0.95\linewidth]{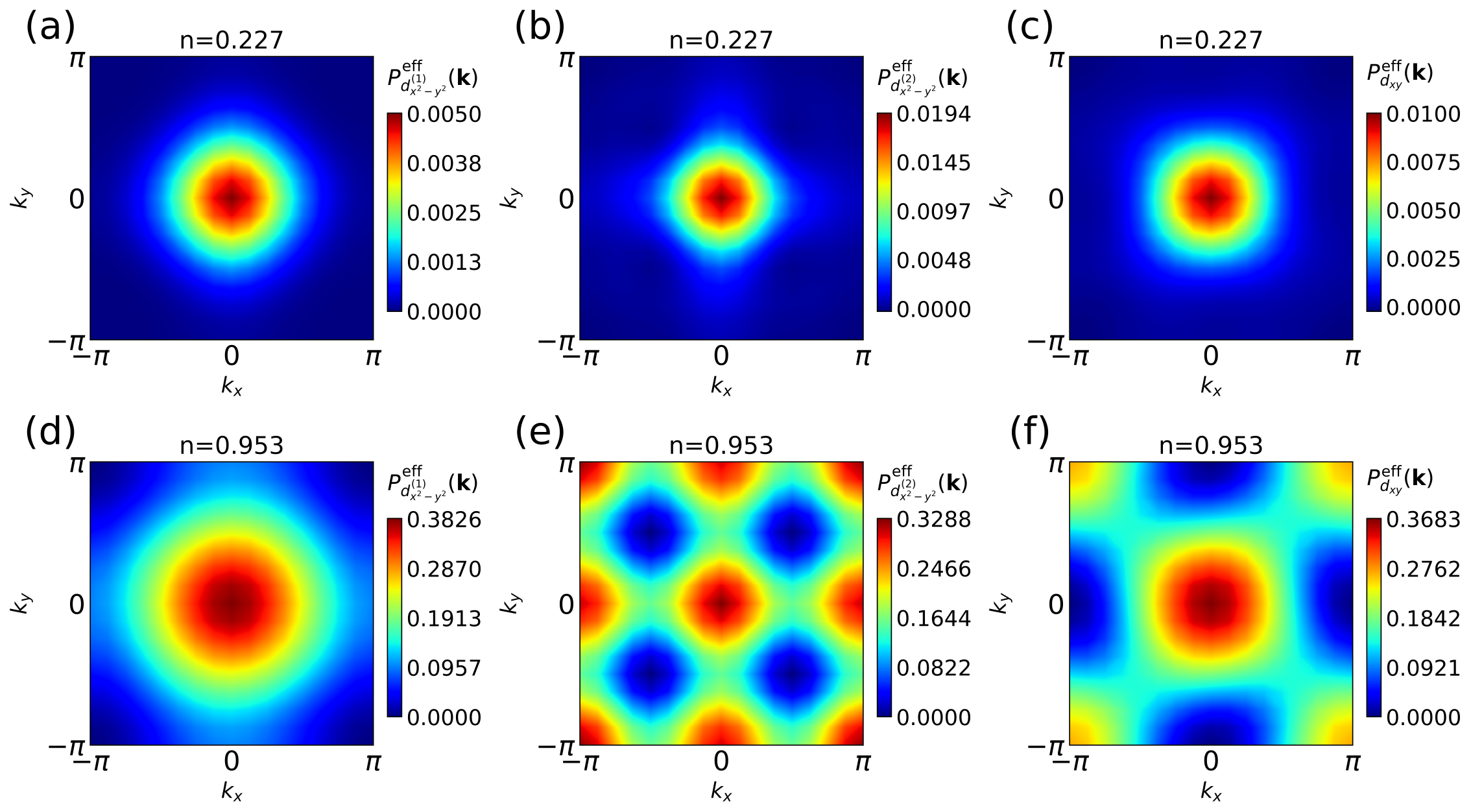}
    \caption{(Color online) Momentum-resolved two-boson correlator $P^{\text{eff}}_\zeta(\mathbf{k})$ for three pairing channels at fixed anisotropy $\alpha = 0.05$, $t^{\prime} = 0.8$, and two representative fillings. 
    (a--c) $P^{\text{eff}}_{d^{(1)}_{x^2 - y^2}}(\mathbf{k})$, $P^{\text{eff}}_{d^{(2)}_{x^2 - y^2}}(\mathbf{k})$, and $P^{\text{eff}}_{d_{xy}}(\mathbf{k})$ at $n = 0.227$. 
    (d--f) Corresponding data at $n = 0.953$. 
    At low filling, the third-nearest-neighbor $d^{(2)}_{x^2 - y^2}$-wave shows the strongest peak, while at high filling, the nearest-neighbor $d^{(1)}_{x^2 - y^2}$-wave component becomes dominant. The $d_{xy}$-wave contribution remains competitive in both cases, highlighting the filling-dependent reordering of pairing structure under strong frustration.}
    \label{fig8}
\end{figure}

Importantly, if one integrates over the full momentum distribution $P^{\text{eff}}_\zeta(\mathbf{k})$ instead of considering only the peak value, then the third-nearest-neighbor $d_{x^2 - y^2}^{(2)}$-wave dominates at $t^{\prime}/t = 0.8$. At $t^{\prime}/t = 0.2$, by contrast, the global and peak behavior agree in singling out the $d_{xy}$-wave. This contrast is further illustrated in Fig.~\ref{fig8}(a--c): at low filling $n = 0.271$, the momentum peak of $P^{\text{eff}}_\zeta(\mathbf{k})$ is largest for the $d_{x^2 - y^2}^{(2)}$-wave channel, followed by $d_{xy}$-wave, with $d_{x^2 - y^2}^{(1)}$-wave being the weakest. At higher filling $n = 0.953$ [Fig.~\ref{fig8}(e--f)], however, the structure reverses—$d_{x^2 - y^2}^{(1)}$-wave becomes dominant, followed by $d_{xy}$-wave and then $d_{x^2 - y^2}^{(2)}$-wave.

To further visualize these trends in real space, we examine the short-range correlator $C^{\text{eff}}_\zeta(i)$ at representative fillings, indicated by the vertical dashed lines in Figs.~\ref{fig7}(a) and ~\ref{fig7}(b). For $t^{\prime}/t = 0.2$, results at $n = 0.656$ and $n = 0.953$ [Figs.~\ref{fig7}(c) and ~\ref{fig7}(d)] show that the $d_{x^2 - y^2}^{(2)}$-wave and $s$-wave exhibit the strongest short-range amplitudes, while $d_{x^2 - y^2}^{(1)}$-wave and $d_{xy}$-wave remain subleading. For $t^{\prime}/t = 0.8$, panels (e) and (f) show the corresponding behavior at $n = 0.321$ and $n = 0.953$, respectively. At low filling, all three $d$-wave components have comparable strengths; at higher $n$, $d_{x^2 - y^2}^{(1)}$-wave and $d_{xy}$-wave dominate in agreement with momentum-space trends. In all cases, the correlators decay rapidly with distance, confirming the short-range nature of bosonic pairing in the CPBM phase.

These observations are consistent with previous findings~\cite{Cao2025}, which suggested that anisotropic third-nearest-neighbor hopping $t^{\prime}$ can enhance $d_{x^2 - y^2}$-wave-like correlations in the Bose-metal regime. Our results support this scenario: at $t^{\prime}/t = 0.2$, both third- and nearest-neighbor $d$-wave  grow significantly with increasing filling, while the $d_{xy}$-wave remains dominant. When $t^{\prime}/t = 0.8$, the dominant channel shifts from $d_{x^2 - y^2}^{(2)}$-wave at low filling to $d_{x^2 - y^2}^{(1)}$-wave at high filling, reflecting a filling-dependent rearrangement of pairing symmetry under strong frustration.

In summary, our symmetry-resolved analysis reveals that the internal pairing structure of the CPBM phase is governed by short-range $d$-wave correlations, with dominant contributions from either $d_{xy}$-wave or $d_{x^2 - y^2}$-wave channels depending sensitively on the degree of frustration and carrier filling. At moderate frustration ($t^{\prime}/t = 0.2$), the $d_{xy}$-wave symmetry consistently dominates across all fillings. In contrast, for stronger frustration ($t^{\prime}/t = 0.8$), the dominant pairing symmetry exhibits a filling-dependent crossover—from third-nearest-neighbor $d_{x^2 - y^2}^{(2)}$-wave at low fillings to nearest-neighbor $d_{x^2 - y^2}^{(1)}$-wave and $d_{xy}$-wave components at higher fillings—revealing a more intricate internal structure. The suppression of $s$-wave pairing and the rapid real-space decay of all pairing channels confirm the non-superfluid and unconventional nature of the CPBM phase. These findings support a microscopic scenario in which the Bose surface emerges from strongly correlated finite-momentum pair scattering between anisotropic orbitals, stabilized by both next-nearest-neighbor hopping and spin anisotropy. The absence of long-range order, combined with a persistent structure of extended $d$-wave correlations, highlights the robustness and tunability of this exotic metallic state, indicating a robust and uniform $d$-wave Bose metal regime~\cite{Feiguin2011,Jiang2013,Cao2024,PhysRevB.78.054520}.

\section{Discussion and Conclusion}
\label{Discussion and Conclusion}

Using large-scale CPQMC simulations, we have systematically explored the stability and evolution of the CPBM phase in the spin-anisotropic attractive Hubbard model. Our results demonstrate that carrier filling $n$, spin anisotropy $\alpha$, and NNN hopping $t^{\prime}$ act cooperatively to stabilize this non-superfluid bosonic metallic phase.

A moderate level of frustration, $t^{\prime}/t \simeq 0.2$, generates a dome-like CPBM region in the $(n,t^{\prime})$ plane centered near $n \approx 0.90$. Under these conditions, the CPBM phase remains robust up to half-filling and even coexists with a weak CDW for $n > 0.95$. This coexistence is absent in the $t^{\prime}=0$ case, where CDW order dominates the high-filling regime~\cite{Cao2024}. These findings underscore the role of $t^{\prime}$ in frustrating real-space nesting, expanding the CPBM region, and suppressing the CDW correlations—refining previous conclusions drawn from nearest-neighbor-only models.

In addition to extending the CPBM regime, $t^{\prime}$ qualitatively reshapes the single-particle FS, inducing both a topology change and a geometric rotation. As $t^{\prime}$ increases, the criteria for the FS to transition from an open to a closed topology become more stringent, and the FS simultaneously undergoes a geometric rotation away from the crystal axes. Our simulations show that the orientation of the resulting BS also evolves accordingly: when the FS rotation is small, the BS remains approximately aligned with the lattice directions, but as the FS becomes more distorted, the BS gradually co-rotates with it. This continuous evolution reveals a geometric coupling between fermionic and bosonic degrees of freedom, and highlights that the stability of the CPBM phase is governed more by FS anisotropy than by FS topology—contrary to earlier classifications of $d$-wave Bose metals that emphasized open versus closed FS contours, such as the distinction between the $d$-wave local Bose liquid (DLBL) and the $d$-wave Bose liquid (DBL)~\cite{PhysRevB.75.235116}.

We further characterize the internal pairing structure of the CPBM phase by analyzing the structure of bosonic pairing channels. At weak frustration ($t^{\prime}/t = 0.2$), the $d_{xy}$-wave component dominates across the entire filling range, and this $d_{xy}$-wave-dominated structure remains largely insensitive to the electron density. However, as the frustration strength increases, the dominant pairing symmetry becomes strongly dependent on filling: at low $n$, the third-nearest-neighbor $d_{x^2 - y^2}^{(2)}$-wave is favored, whereas at high $n$, the nearest-neighbor $d_{x^2 - y^2}^{(1)}$-wave component takes over. This trend refines earlier findings~\cite{Cao2025}, which suggested that around $n \approx 0.65$, increasing frustration generally drives a transition in the dominant pairing channel from $d_{xy}$-wave to $d_{x^2 - y^2}^{(1)}$-wave. Our results reveal that this transition also depends sensitively on the filling $n$—in particular, even at large $t^{\prime}$, the $d_{x^2 - y^2}^{(1)}$-wave does not dominate uniformly across all fillings. Instead, the CPBM phase exhibits a nontrivial pairing structure that evolves with both carrier density and frustration strength. The leading pairing symmetry shifts in a filling-dependent manner, highlighting a richer internal structure. This refined picture further differentiates the CPBM from conventional superconducting or density-ordered phases, providing a clearer microscopic signature of bosonic metallicity.~\cite{Cao2025,Cao2024}.

Taken together, these results elucidate that variations in the   particle filling $n$ and NNN hopping $t^{\prime}$ in the $t-t^{\prime}-U$ Hubbard model significantly affect the properties of the Bose metallic phase. They further clarify how geometric frustration and band-structure deformation enhance the stability and internal coherence of the CPBM. These findings offer valuable guidance for exploring such exotic phases within experimentally accessible parameter regimes, particularly in ultracold atomic gases with spin-dependent optical lattices~\cite{Serwane2011Deterministic,Jau2016Entangling,Zeiher2016ManyBody}, or in solid-state systems exhibiting altermagnetic correlations~\cite{Smejkal2022Emerging,Smejkal2022Beyond,li2025enhancementdwavepairingstrongly}. Characteristic signatures—including a dome-shaped CPBM region, suppressed CDW order, and co-rotation of Fermi and Bose surfaces—serve as clear indicators and conceptual guides for future investigations of non-superfluid bosonic metals.

\section*{Acknowledgments}

This work is supported by the National Natural Science Foundation of China~(Grant No. 12204130), Shenzhen Start-Up Research Funds~(Grant No. HA11409065), Shenzhen Key Laboratory of Advanced Functional CarbonMaterials Research and Comprehensive Application (Grant No. ZDSYS20220527171407017). T.Y. acknowledges supports from Natural Science Foundation of Heilongjiang Province~(No.~YQ2023A004).

\appendix
\section{Lattice size effect}
\label{appendix:S1}
\begin{figure}[tb!]
    \includegraphics[width=\linewidth]{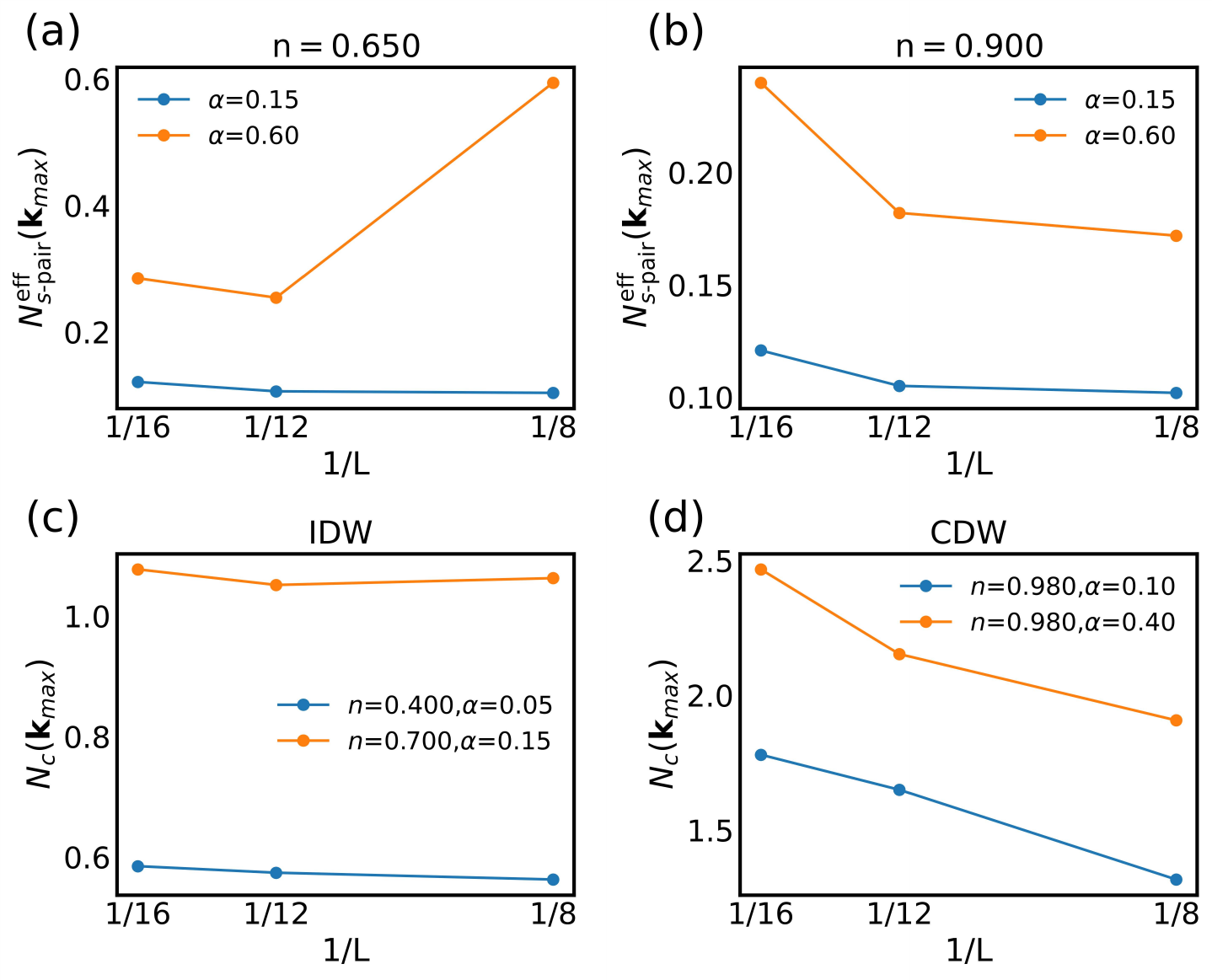}
    \caption{(Color online) Finite-size scaling of pairing and charge correlations for representative parameter sets corresponding to four distinct quantum phases.(a) Effective pair structure factor $N^{\text{eff}}_{s\text{-pair}}(\mathbf{k}_{\text{max}})$ at $n = 0.650$ for $\alpha = 0.15$ (CPBM) and $\alpha = 0.60$ (s-SF), plotted versus $1/L$ for $L = 8$, 12, 16. The CPBM signal remains nearly flat, indicating short-range pairing, while the s-SF signal increases with system size, consistent with long-range coherence.(b) Same analysis at $n = 0.900$, showing consistent behavior: $N^{\text{eff}}_{s\text{-pair}}(\mathbf{k}_{\text{max}})$ remains size-independent for CPBM, but grows for s-SF.(c) Charge structure factor $N_c(\mathbf{k}_{\text{max}})$ for IDW phases at $(n, \alpha) = (0.400, 0.05)$ and $(0.700, 0.15)$. In both cases, the peak values remain robust across increasing lattice size, indicating stable incommensurate charge modulation.(d) $N_c(\mathbf{k}_{\text{max}})$ for the CDW phase at $n = 0.980$, comparing $\alpha = 0.10$ and $\alpha = 0.40$. The peak height increases with system size, confirming that CDW order persists in the thermodynamic limit.}
    \label{S1}
\end{figure}

To verify that the phases identified in main text are not artifacts of finite system size, we perform a finite-size scaling analysis using $L \times L$ lattices with $L = 8, 12, 16$ at $t^{\prime} = 0.8$. Fig.~\ref{S1} displays the behavior of the dominant momentum peaks in the pairing and charge structure factors across representative points in the CPBM, s-SF, IDW, and CDW regimes.
Figs.~\ref{S1}(a) and \ref{S1}(b) present the evolution of $N^{\text{eff}}_{s\text{-pair}}(\mathbf{k}_{\text{max}})$ at $n = 0.650$ and $n = 0.900$, respectively. For $\alpha = 0.15$, corresponding to the CPBM phase, the signal remains nearly constant with increasing $L$, consistent with a short-range paired metallic state. By contrast, for $\alpha = 0.60$, the pairing strength increases markedly with system size, indicating the emergence of long-range phase coherence in the s-SF phase. These scaling trends offer an independent confirmation of the CPBM/$s$-SF distinction established from momentum and real-space diagnostics. In Fig.~\ref{S1}(c), we analyze $N_c(\mathbf{k}_{\text{max}})$ for the IDW phase at $(n, \alpha) = (0.400, 0.05)$ and $(0.700, 0.15)$. The charge structure factor remains nearly size-independent in both cases, supporting the persistence of incommensurate charge order in the thermodynamic limit. Finally, Fig.~\ref{S1}(d) shows the CDW regime at $n = 0.980$, comparing $\alpha = 0.10$ and $\alpha = 0.40$. In both cases, the charge peak at $\mathbf{k} = (\pi, \pi)$ grows with increasing system size, confirming that commensurate charge density wave order not only survives, but is enhanced as $L \rightarrow \infty$.

Together, these results validate the robustness of all four phases—CPBM, s-SF, IDW, and CDW—under finite-size scaling. The contrasting behavior of the pairing and charge structure factors serves as a reliable diagnostic of the underlying order, and further supports the phase boundaries established in earlier sections.
\\
\\
\\
\\
\\
\\
\\
\\
\\
\\
\\
\\
\\
\bibliography{references}
\end{document}